\documentclass[twocolumn]{aastex61}

\usepackage{gensymb}
\usepackage{amsmath}

\newcommand{\beq}{\begin{equation}}
\newcommand{\eeq}{\end{equation}}

\newcommand{\hi}{H{\sc i}~}
\newcommand{\HI}{H{\sc i}}
\newcommand{\kms}{km ${\rm s^{-1}}$~}
\newcommand{\kmsa}{km ${\rm s^{-1}}$}
\newcommand{\thi}{$t_{H{\sc I}}~$}
\newcommand{\tHI}{$t_{H{\sc I}}$}

\shorttitle{A new probe of line-of-sight magnetic field tangling}
\shortauthors{Clark}

\begin{document}

\title{A new probe of line-of-sight magnetic field tangling}

\correspondingauthor{S. E. Clark}
\email{seclark@ias.edu}

\author[0000-0002-7633-3376]{S. E. Clark}
\altaffiliation{Hubble Fellow}
\affil{Institute for Advanced Study,
1 Einstein Drive,
Princeton, NJ 08540}



\begin{abstract}

The Galactic neutral hydrogen (\HI) sky at high Galactic latitudes is suffused with linear structure. Particularly prominent in narrow spectral intervals, these linear \hi features are well aligned with the plane-of-sky magnetic field orientation as measured with optical starlight polarization and polarized thermal dust emission. We analyze the coherence of the orientation of these features with respect to line-of-sight velocity, and propose a new metric to quantify this \hi coherence. We show that \hi coherence is linearly correlated with the polarization fraction of 353 GHz dust emission. \hi coherence constitutes a novel method for measuring the degree of magnetic field tangling along the line of sight in the diffuse interstellar medium. We propose applications of this property for \HI-based models of the polarized dust emission in diffuse regions, and for studies of frequency decorrelation in the polarized dust foreground to the cosmic microwave background (CMB). 

\end{abstract}




\section{Introduction} \label{sec:intro}

Magnetic fields are notoriously difficult to measure in the interstellar medium (ISM). Observational tracers of magnetic fields include starlight polarization, polarized thermal dust emission, the Zeeman effect, Faraday rotation of linear polarization, and synchrotron emission \citep[e.g.,][]{Ferriere:2001}. Each of these probes different components of the magnetic field vector, in different phases of the ISM, making a complete description of the magnetic field structure in any region a formidable challenge. Consequently, many questions remain about the structure of the Galactic magnetic field \citep[e.g.][]{Beck:1996, Haverkorn:2015}, although significant progress has been made recently by simultaneously constraining multiple components of the field \citep[e.g.][]{Jaffe:2010, Jansson:2012}.

Recently it was shown that the diffuse neutral hydrogen (\HI) throughout the high Galactic latitude sky is populated by thin, linear features, sometimes referred to as \hi fibers \citep{Clark:2014}. These linear \hi structures are extremely well aligned with the plane-of-sky magnetic field as probed by both starlight polarization \citep{Clark:2014} and polarized dust emission \citep{Clark:2015, Kalberla:2016}. The structure of the diffuse neutral ISM is deeply coupled to the ambient magnetic field, and the morphology of \hi can be used to study magnetism in diffuse environments. 

Several specific applications of this phenomenon for measuring interstellar magnetic fields have been proposed. In regions dominated by a strong mean magnetic field, \hi orientation can be used to measure the magnetic field strength via a modified Chandrasekhar-Fermi method \citep{Chandrasekhar:1953}, as demonstrated with observations of linear structures in \hi absorption \citep{Clark:2014}. At high Galactic latitudes, \hi orientation can be used to constrain the polarized dust foreground to cosmology experiments \citep{Clark:2015}. These insights have been used to construct models of the Galactic polarized dust emission based in part on the orientation of linear \hi structures \citep{Ghosh:2017}. In this Letter, we demonstrate a new use of \hi morphology and velocity structure: to trace the degree of interstellar magnetic field tangling along the line of sight. 

Measuring the degree of magnetic disorder is of interest both for our understanding of the diffuse ISM and for the search for inflationary gravitational wave $B$-mode polarization in the cosmic microwave background (CMB). These $B$-modes, if detected, would be definitive evidence for inflation \citep{Seljak:1997}, but are obscured by partially polarized thermal emission from magnetically aligned dust grains in the ISM \citep{Flauger:2014, PlanckBicep:2015}. Current foreground subtraction techniques rely on measuring the polarized dust signal toward the peak of the dust spectral energy distribution (SED), and extrapolating its contribution to the target frequency. Polarized dust emission is an integral quantity: the vector sum of polarized intensity from magnetically aligned dust grains along the line of sight. Line-of-sight variation in the magnetic field orientation and dust emission properties thus gives rise to ``frequency decorrelation," wherein the polarized sky at one frequency is not simply a scaled version of the polarized sky at another frequency. Observational constraints on this effect are still being refined, after initial measurements of significant decorrelation in the \textit{Planck} data between 353 GHz and 217 GHz \citep{Planck:L} were found to not be statistically meaningful \citep{Sheehy:2017, Planck:LIV}. Frequency decorrelation must be present at some level, and incorrect assumptions about the frequency dependence of polarized foregrounds can upwardly bias measurements of the tensor-to-scalar ratio. In this Letter we demonstrate a novel estimator of one of the primary physical causes of frequency decorrelation: differently oriented magnetic fields along the line of sight. We propose a new use of \hi morphology for modeling polarized dust foregrounds. 

\section{Methods}

We use \hi data from the Galactic Arecibo L-Band Feed Array Survey \citep[GALFA-\HI;][]{Peek:2017}, a survey of the 21-cm \hi line over the Arecibo sky (decl. $-1\degree17'$ to $+37\degree57'$ across all R.A.) at $4'$ spatial resolution and $0.18$ \kms spectral resolution, with $150$ mK median rms noise per 1 \kms channel. We analyze a $\sim 3417$ deg$^2$ region of GALFA-\hi data at high Galactic latitudes, avoiding the edges of the Arecibo declination range where telescope scan artifacts are more prominent. We quantify the orientation of linear structures in the GALFA-\hi survey using the Rolling Hough Transform \citep[RHT;][]{Clark:2014}. The RHT is a machine vision algorithm that measures linear power as a function of orientation angle on the plane of the sky. Given image-plane data $I(x, y)$ the RHT returns linear intensity $R(\theta, x, y)$. We run the RHT over the entire GALFA-\hi sky, on velocity channels of width $\delta v = 3$ \kmsa, from $v_{lsr}$ = -36.4 \kms to $v_{lsr}$ = +36.4 \kms (where $v_{lsr}$ is the velocity with respect to the local standard of rest). The RHT parameters are the same as those used in \citet{Clark:2015}. We also make use of a stray radiation-corrected column density map created from the velocity-integrated \hi brightness temperature under the optically thin assumption.  The \hi data, including the column density map, and the RHT data product are all publicly available as part of the GALFA-\hi Data Release 2 \citep{Peek:2017}. 

\begin{figure*}
\centering
\includegraphics[trim={1cm 1cm 0 0},width=\textwidth]{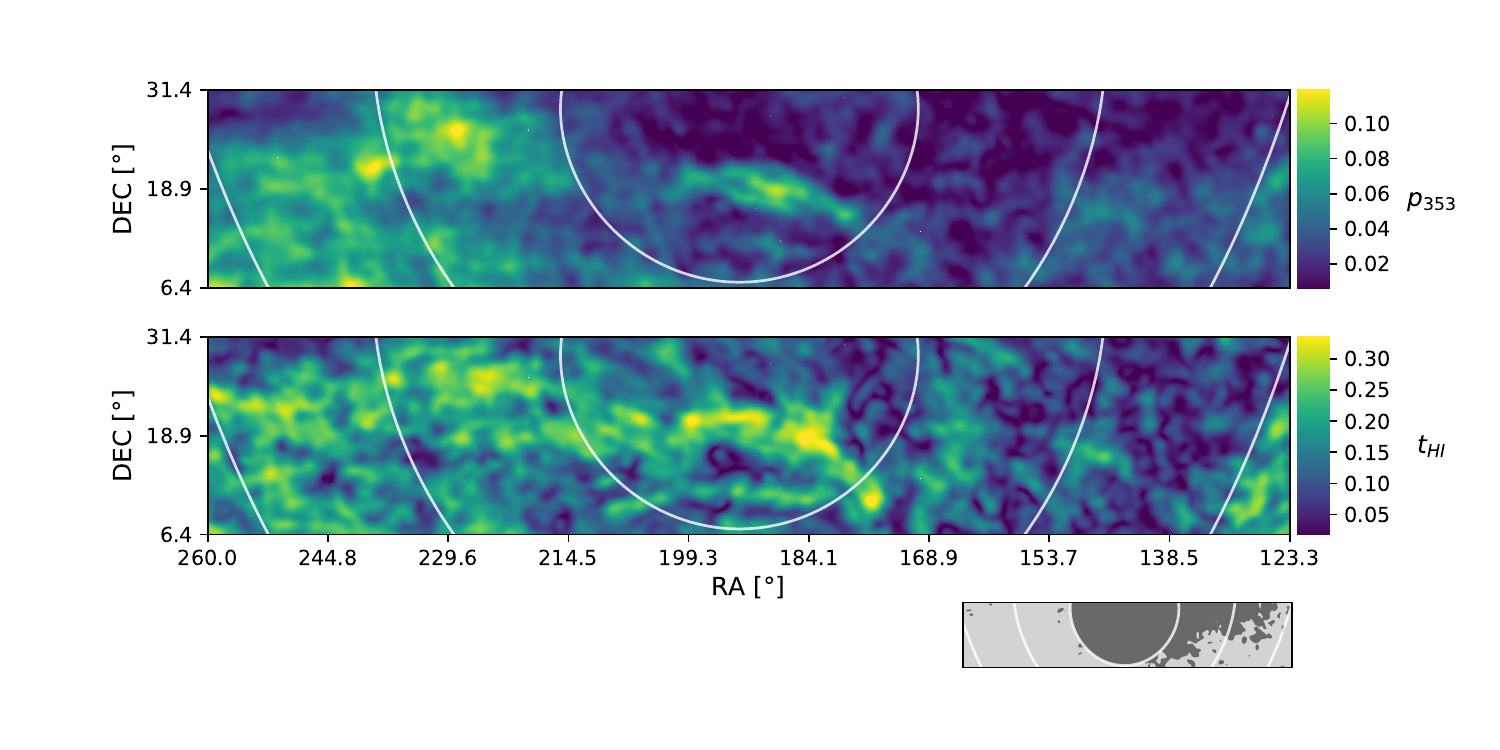}
\caption{Sky maps of the \textit{Planck} 353 GHz polarization fraction (top) and \hi coherence (bottom). Note the different ranges on the color bars, which linearly map the 5-95\% intensity range of their respective quantities. Both $p_{353}$ and \thi are shown at an angular resolution of $90'$. White lines are lines of constant Galactic latitude: $b = 30\degree, 50 \degree, 70\degree$, from left to center. The small inset shows the same region of sky, where regions in dark gray are masked from the analysis in Figures \ref{fig:hist2d} and \ref{fig:hist2dnhip353}. }\label{fig:skyplot}
\end{figure*}

For each velocity channel, we calculate the point estimate of the \hi orientation. That is, we define

\begin{align}
Q_{RHT}(x, y, v) = \int \mathrm{cos} \left(2 \theta \right) \cdot R\left(\theta, x, y, v \right) \, \mathrm{d}\theta\\
U_{RHT}(x, y, v) = \int \mathrm{sin} \left(2 \theta \right) \cdot R\left(\theta, x, y, v \right) \, \mathrm{d}\theta,
\end{align}

such that each pixel in each velocity channel has a value

\beq
\theta_{RHT} = \frac{1}{2} \mathrm{arctan} \frac{U_{RHT}}{Q_{RHT}},
\eeq

measuring the local orientation of linear \hi features. We mask any pixels that have zero RHT-measured linear intensity, i.e., any ($x_0$, $y_0$) where $\int R\left(\theta, x_0, y_0 \right) \, \mathrm{d}\theta = 0$.

We define a metric for the degree of coherence of \hi orientation as a function of velocity. This parameter, which we call $t_{HI}$, is constructed analogously to the dust polarization fraction $p$, and similarly has a theoretical range of [0, 1]. We define new Stokes-like parameters using the orientation of \hi weighted by the local \hi intensity, and integrate over the line-of-sight velocity ($v_{lsr}$). That is,

\begin{align}
Q_{H{\sc I}} = \int I_v \, \mathrm{cos} \left(2 \theta_{RHT} (v) \right) \, \mathrm{d}v \\
U_{H{\sc I}} = \int I_v \, \mathrm{sin} \left(2 \theta_{RHT} (v) \right) \, \mathrm{d}v,
\end{align}

where $I_v$ is the \hi intensity in a given velocity channel.

Our measure of how much the orientation of \hi varies along the line of sight is then 

\beq
\label{eq:hi_coherence}
t_{H{\sc I}} = (Q_{H{\sc I}}^2 + U_{H{\sc I}}^2)^{1/2} / I_{H{\sc I}},
\eeq

where $I_{H{\sc I}}$ represents the total \hi intensity integrated over the same velocity range. When $\theta_{RHT}$ varies significantly along the line of sight, $t_{H{\sc I}}$ will have a lower value. By contrast, when the orientation of \hi is fairly uniform in velocity space, or when only a single linear \hi feature dominates the \hi emission along the line of sight, the magnitude of $t_{H{\sc I}}$ will be higher. We refer to \thi as ``\hi coherence" to evoke this behavior.

We compare \hi coherence with the measured fractional polarization of the \textit{Planck} 353 GHz emission. This is defined from the linear Stokes parameters as

\beq
\label{eq:pfrac}
p_{353} = P_{353}/I_{353}.
\eeq

The polarized intensity $P_{353}$ is naively equal to $({Q_{353}^2 + U_{353}^2})^{1/2}$, and so Equation \ref{eq:hi_coherence} is defined in analogy to Equation \ref{eq:pfrac}. In practice, however, the quadratic dependence on $Q$ and $U$ makes $P_{353}$ a noise-biased quantity. To mitigate this we compute the debiased point estimate $\hat{P}_{353}$ using the \citet{Plaszczynski:2014} modified asymptotic estimator \citep[see also][]{Montier:2015a, Montier:2015b}. $\hat{P}_{353}$ is computed from the publicly available \textit{Planck} 353 GHz Stokes parameters and covariance matrices \citep{Planck:2016}. For clarity here we indicate debiased quantities with a hat, but $p_{353}$ is henceforth always constructed from the debiased $\hat{P}_{353}$.

The values $t_{H{\sc I}}$ and $p_{353}$ are compared at an angular resolution of 90$'$, computed by convolving $Q_{RHT}$, $U_{RHT}$, ${Q}_{353}$, and ${U}_{353}$ with a Gaussian kernel with FWHM= 90$'$. We also smooth the 353 GHz noise covariance matrix prior to computing $p_{353}$ \citep[see Appendix A of][]{PlanckXIX}. Any pixels that have a signal-to-noise ratio (SNR) $<$ 2 on $p_{353}$ at the $90\%$ confidence level are masked from the analysis. The region of sky analyzed in Section \ref{sec:results} has a median $p_{353}/\sigma_{p}\sim 11$ at $90'$, where $\sigma_p$ is the standard deviation of the \citet{Plaszczynski:2014} estimator.

\section{Results}\label{sec:results}

We find that \thi and $p_{353}$ are correlated across this region of high Galactic latitude sky. Figure \ref{fig:skyplot} shows a heatmap of both quantities at $90'$ resolution on the plane of the sky. In the region shown, the GALFA-\hi $\left| {v_{lsr}} \right| \le 90$ \kms column density map smoothed to an angular resolution FWHM=$90'$ has a maximum value of $N_{HI} \sim 9.3 \times 10^{20}$ $\mathrm{cm}^{-2}$, and a median value of $N_{HI} \sim 2.4 \times 10^{20}$ $\mathrm{cm}^{-2}$. At these low column densities, the scatter in $p_{353}$ for a given $N_{HI}$ is large, and $p_{353}$ is not strongly dependent on $N_{HI}$ \citep{PlanckXIX}.

The range of values of \thi in the region analyzed is larger than that of $p_{353}$. Although \thi is not a physical model of $p_{353}$, the larger values of \thi can be understood phenomenologically. \hi coherence quantifies the velocity-space alignment of linear \hi structures \textit{on the plane of the sky}. The true degree of line-of-sight magnetic field tangling, however, depends on the three-dimensional magnetic field orientation. Simple models for polarized dust emission compute linear Stokes parameters that are proportional to a density-weighted integral of the local magnetic field orientation, so that

\beq
Q_{dust} \propto \int \rho \, \mathrm{cos} \left(2 \theta \right) \mathrm{cos}^2 \gamma \, d\mathrm{s}
\eeq
\beq
U_{dust} \propto \int \rho \, \mathrm{sin} \left(2 \theta \right) \mathrm{cos}^2 \gamma \, d\mathrm{s},
\eeq

where $\rho$ is the local dust density, $\theta$ is the plane-of-sky magnetic field orientation, $\gamma$ is the angle between the line of sight and the local magnetic field, and $s$ is distance \citep[e.g.,][]{Wardle:1990}. The construction of \thi thus mimics this model, with local \hi intensity replacing local dust intensity, $\theta_{RHT}$ replacing the plane-of-sky magnetic field orientation, and $v_{lsr}$ replacing distance. Missing from the \thi construction, however, is information about the relative orientation between the line of sight and the local magnetic field orientation (the $\mathrm{cos}^2 \gamma$ term). Loosening this implicit assumption that the magnetic fields contain a negligible line-of-sight component would lower the values of \tHI. Incorporating a model of $\mathrm{cos}^2 \gamma$ into a \tHI-like quantity might reduce some of the scatter in the \thi - $p_{353}$ relationship; alternatively, \thi and $p_{353}$ could be used to fit an estimate of $\gamma$.

\begin{figure}[h!]
\centering
\includegraphics[width=\columnwidth]{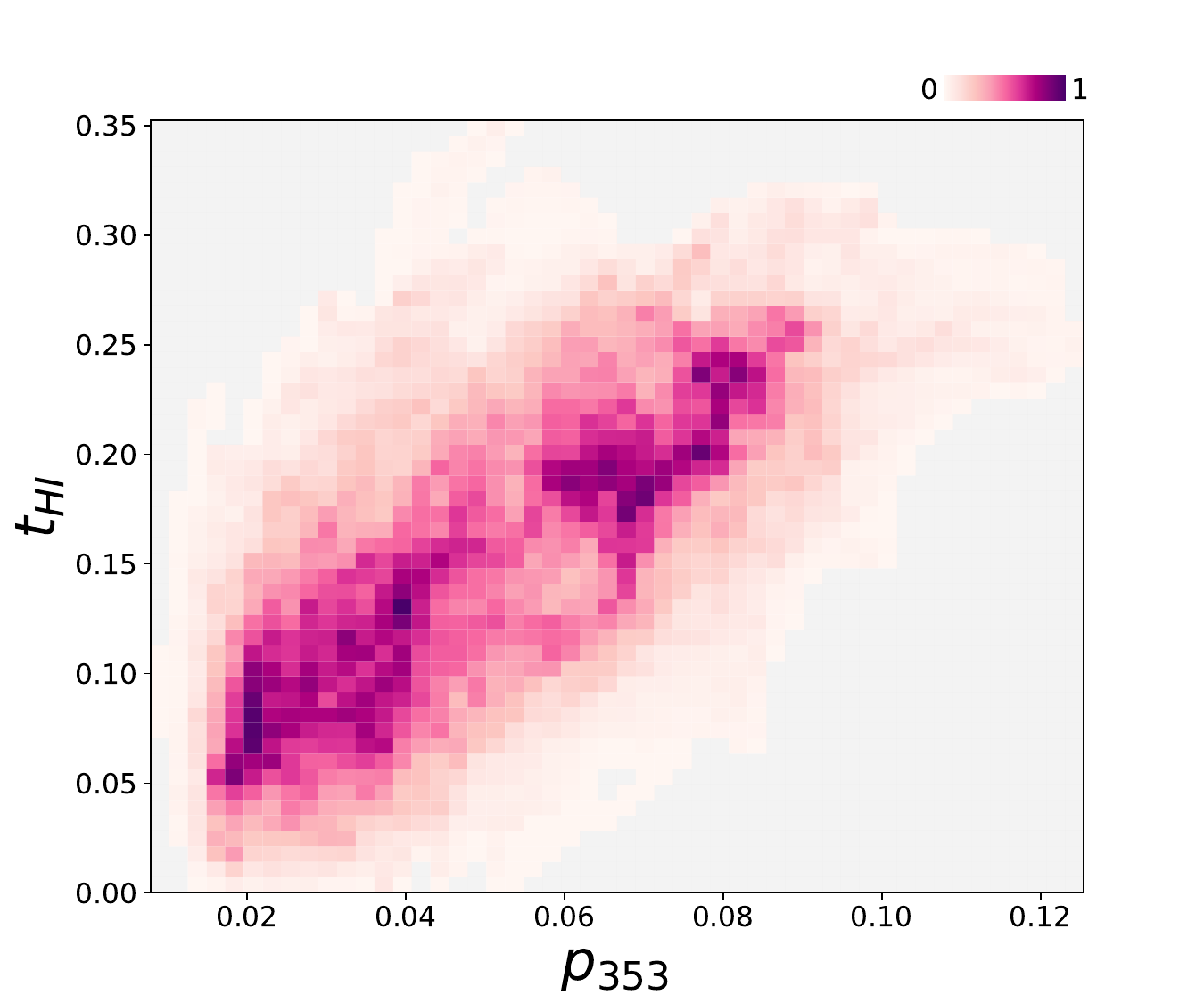}
\caption{Two-dimensional histogram of the \textit{Planck} 353 GHz polarization fraction ($p_{353}$) and \hi coherence (\tHI). Colors represent the (normalized) pixel count value on a linear scale. Gray grid cells represent regions of the histogram with zero power. The region of sky analyzed is that in Figure \ref{fig:skyplot}, excluding pixels at Galactic latitudes $b > 70$ and pixels where the velocity range analyzed accounts for less than 90\% of the $\pm 90$ \kms \hi column, as discussed in the text. }\label{fig:hist2d}
\end{figure}

\begin{figure}[h!]
\centering
\includegraphics[width=\columnwidth]{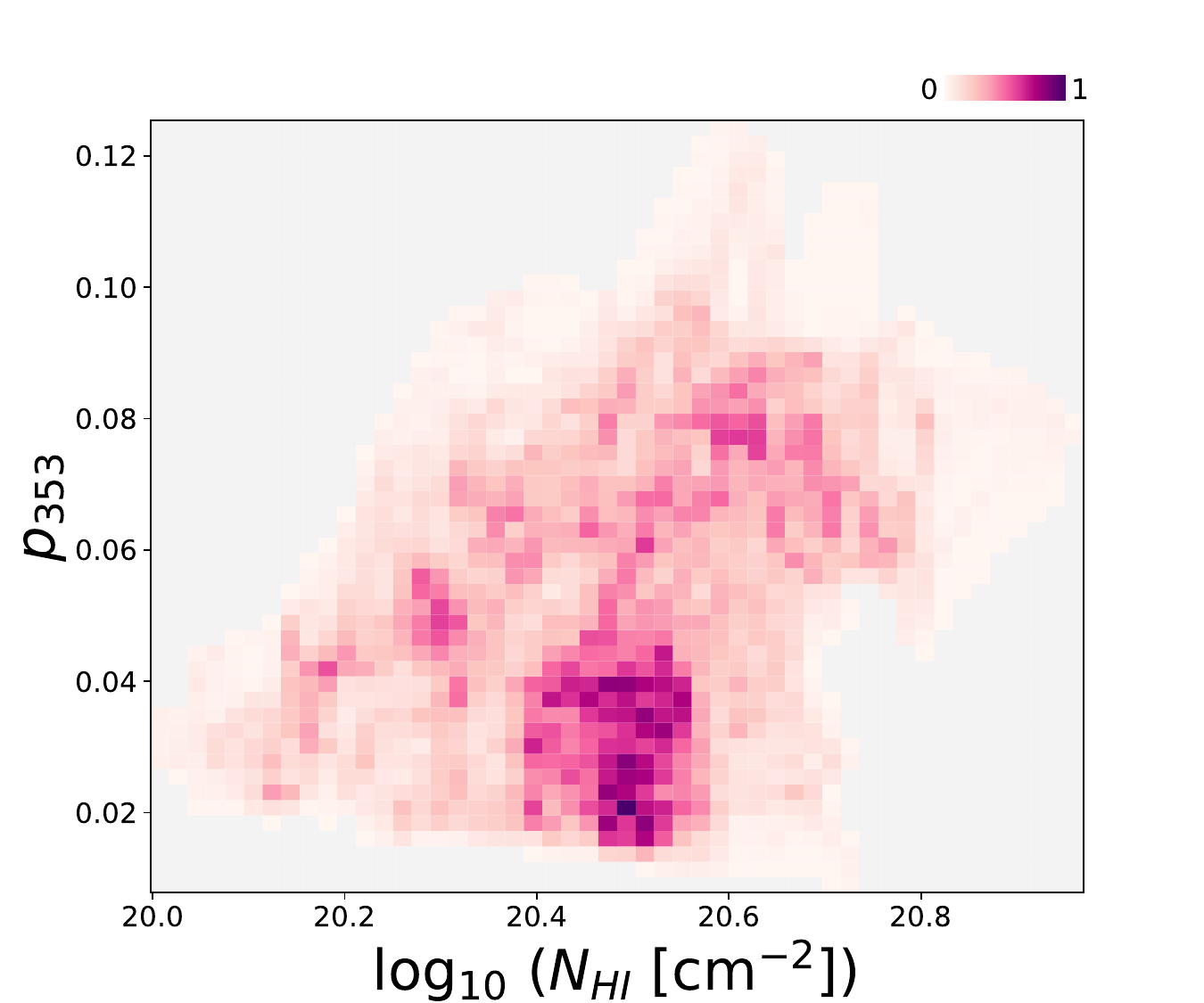}
\caption{As in Figure \ref{fig:hist2d}, but for $p_{353}$ and $\mathrm{log}_{10}$ of the stray radiation-corrected GALFA-\hi column density integrated over $\pm 90$ \kmsa. The region of sky analyzed is the same as in Figure \ref{fig:hist2d}, and quantities are compared at FWHM=$90'$. }\label{fig:hist2dnhip353}
\end{figure}

Figure \ref{fig:hist2d} shows a two-dimensional histogram of $p_{353}$ and \thi for the subset of the data indicated in the inset of in Figure \ref{fig:skyplot}. We consider the data in Figure \ref{fig:skyplot} below a Galactic latitude of $b = 70\degree$, to avoid latitudes near Galactic zenith where uncertainties in the \textit{Planck} data are high and contamination from the cosmic infrared background may be significant \citep{PlanckXIX}. In order to include only sightlines where an appreciable fraction of the total \hi column has been analyzed, we compute the ratio of the GALFA-\hi data integrated over $\left| v_{lsr} \right| \le 36.4$ \kms to the data integrated over $\left| v_{lsr} \right| \le 90$ \kms at $90'$ resolution. We mask pixels where this ratio is less than $0.9$, though results are not strongly sensitive to this choice of ratio. For the masked pixels, more than $10\%$ of the total \hi column apparently lies at velocities outside of the range analyzed with the RHT. The individual velocity channel maps in the GALFA-\hi data are not corrected for stray radiation, and so there also exist pixels where this ratio is greater than unity, due to noise. We retain these pixels but note that their removal does not qualitatively change Figure \ref{fig:hist2d}. Figure \ref{fig:hist2d} shows a clear correlation between $p_{353}$ and \tHI. A simple linear least-squares fit to this data yields \thi$ = 2.1 p_{353} +  0.049$, with a correlation coefficient $r\sim0.7$. 

The corresponding two-dimensional histogram between $\mathrm{log}_{10} N_{HI}$ and $p_{353}$ is shown in Figure \ref{fig:hist2dnhip353}. In this low column density ($< 10^{21}$ $\mathrm{cm}^{-2}$) region there is no indication of the anticorrelation between $N_{H}$ and $p_{353}$ found at higher column densities \citep{PlanckXIX}, and in fact this region displays a weak but significant positive correlation, with a linear correlation coefficient $r\sim0.4$. These results do not depend strongly on the cuts made to the data. To estimate the robustness of these correlations we performed a block bootstrap analysis by dividing the sky into $100' \times 100'$ regions, sampling these regions with replacement $10^4$ times, and fitting each sample with a least-squares linear regression. We find a $95\%$ confidence interval for $r$ of [0.65, 0.71] for the $p_{353}$-\thi relation and [0.33, 0.41] for the $N_{HI}$-$p_{353}$ relation. We conclude that \hi coherence is positively correlated with $p_{353}$ in low column regions where $N_{HI}$ is not as strongly predictive of $p_{353}$.

\section{Discussion}

This work extends our understanding of the link between neutral hydrogen morphology and the interstellar magnetic field. We establish for the first time that the velocity-space variation of \hi orientation is a probe of line-of-sight magnetic field tangling. This is important both for our understanding of the magnetized interstellar medium and potentially for measurements of the polarized foreground to cosmology experiments. 

The polarized dust emission at high Galactic latitudes is variable on the plane of the sky. \citet{PlanckXIX} found that spatial variation in the fractional dust polarization at $353$ GHz is anti-correlated with variation in the polarization angle dispersion -- that is, regions with low fractional polarization correspond to regions where the projected polarization angle varies rapidly on the plane of the sky. This suggests that depolarization of the integrated dust emission is primarily caused by disordered magnetic fields on scales smaller than the polarization angle dispersion is measured ($\sim 1\degree$). It seems likely that magnetic field tangling is primarily responsible for the spatial variation of $p_{353}$, rather than, for example, variable grain alignment efficiency \citep[see also][]{PlanckXX}. The present work supports this conclusion by demonstrating that $p_{353}$ is correlated with the frequency-space variation in \hi orientation, a quantity derived from data (\HI) that is independent of the physics of grain alignment. This finding is consistent with expectations from grain alignment theory \citep[e.g.,][]{Lazarian:2007}.

Polarized dust emission is affected by physics at a range of scales, including the Galactic magnetic field structure and turbulence in the diffuse ISM, as well as dust grain population and alignment properties. 
A phenomenological model of polarized dust emission as the sum of $N$ independent dust layers was shown to reproduce the one-point statistics of $p_{353}$ and $\theta_{353}$ over the southern Galactic cap when $N \sim 4-9$ \citep{PlanckXLIV}. The authors fit a single mean magnetic field orientation and assume the turbulent component of the Galactic magnetic field to be Gaussian and isotropic. This approach was expanded to include constraints from the \textit{Planck} 353 GHz $EE$, $BB$, and $TE$ power spectra by \citet{Vansyngel:2017}. These approaches, while not entirely physical, provide useful prescriptions for generating statistically representative, independent realizations of the polarized sky. By using power spectra constraints, \citet{Vansyngel:2017} effectively introduced the observed correlation between density structures and the ambient magnetic field, which creates a $TE$ correlation and a non-unity $EE/BB$ ratio \citep{Clark:2015, PlanckXXXVIII}. In the \citet{Ghosh:2017} model, this correlation is introduced by a single polarization layer in which the plane-of-sky magnetic field is perfectly aligned with \hi structures associated with the cold neutral medium. 

The work presented in this Letter suggests that new models can be constructed in which \hi binned into velocity intervals serve as separate ``layers," each with \hi structure orientation and \hi intensity providing a template for the local magnetic field orientation and local dust intensity, respectively. This is a data-driven way to incorporate the anisotropy of magnetized turbulence, which is missing from the \citet{PlanckXLIV} approach but a relevant astrophysical constraint \citep[e.g.,][]{Brandenburg:2013}. Indeed, magnetohydrodynamic turbulence predicts that elongated eddies rotate perpendicular to the local magnetic field, in agreement with the observation that structures in thin \hi velocity channels are elongated in the direction of the field \citep{Clark:2014, Lazarian:2018}. 
The \hi velocity layer approach outlined here would allow dust emission prescriptions to be mapped onto physical spatial variation, thereby complementing purely frequency-space dust models \citep[e.g.,][]{Hensley:2017}.

Modeling the polarized dust emission at high Galactic latitude is crucial for characterizing the polarized foreground to the CMB. Because the polarized dust emission peaks at frequencies much higher than the peak CMB emission, current foreground dust subtraction techniques use sky measurements at dust-dominated frequencies  
extrapolated to CMB-dominated frequencies of interest. This extrapolation implicitly assumes that the temperature of dust-emitting regions is uniform along the line of sight. If instead multiple regions (clouds) at different distances along the line of sight have different temperatures, they will each contribute differently to the emission at a given frequency. Measured Stokes parameters will change with frequency as clouds with different polarization properties contribute according to their SEDs. In particular, clouds along the line of sight that experience differently oriented magnetic fields -- thus producing emission that is linearly polarized at different angles -- will significantly decorrelate the polarized dust signal between $\sim$350 and $\sim$150 GHz \citep{Tassis:2015, Poh:2017}. 

If regions of significant misalignment of the magnetic field along the line of sight can be identified, they can be excluded from a CMB polarization analysis. One probe of the degree of magnetic field misalignment along the line of sight requires optical starlight polarization measurements at a number of different distances within a single CMB experiment beam \citep[][]{Tassis:2015}. The synthesis of \textit{Gaia} \citep{Gaia:2016} data for stellar distances and the planned PASIPHAE\footnote{http://pasiphae.science} experiment to measure polarization toward millions of high Galactic latitude stars will enable this sort of magnetic tomography. 

For estimating the frequency decorrelation of polarized dust CMB foregrounds, however, it is necessary to know only what lines of sight harbor significantly misaligned magnetic fields, and not \textit{where} in distance the different magnetic fields lie. Thus, this work demonstrates that the local intensity and orientation of linear \hi features in different frequency intervals can identify regions with potentially problematic line-of-sight magnetic field tangling. Masks can be constructed from maps of \thi or a similar metric. Additionally, \hi morphology and the \hi brightness in different line-of-sight velocity channels can be combined with dust SED prescriptions to model the polarized foreground as discussed above, and used to estimate the expected frequency decorrelation. This work utilizes \hi data in the velocity range $\left| v_{lsr} \right| \le 36.4$ \kmsa. We note that a wider velocity range may be necessary for lines of sight that intercept high-velocity clouds in the Galactic halo, some of which may contain detectable dust emission \citep{MAMD:2005, Peek:2009}.

The large-scale structure of the Galactic magnetic field is a major open question, complicated by the fact that various observational probes trace different components of the magnetic field, in different phases of the ISM. The synthesis of multiwavelength data is therefore necessary to understand both the overall structure of the magnetic field and its distribution between ISM phases. Studies of the magnetic field often distinguish between components of the field, characterized by their isotropy in direction and strength, that originate from different physical processes (dynamos, turbulence, etc.) and contribute differently to observables like the total and polarized synchrotron intensity and the Faraday rotation measure \citep[RM; e.g.,][]{Jaffe:2010}. The relative dominance of these magnetic field components likely changes with ISM environment: recent multiwavelength work on the Galactic plane suggests that the Galactic magnetic field may be more ordered in dust-emitting regions than the average field traced by synchrotron and RM data \citep{Jaffe:2013}. Additional constraints from the velocity coherence of \hi orientation may be useful for disentangling the various components of the Galactic magnetic field in different ISM phases.

This work illustrates a promising avenue for magnetic tomography with \HI. If \hi can be mapped to three dimensions, this will constitute an independent, local probe of the magnetic field orientation. This plane-of-sky information is complementary to tracers of the line-of-sight component of the magnetic field. In particular, the synthesis of Faraday tomography \citep[][]{Burn:1966, Brentjens:2005} and three-dimensional dust maps \citep[e.g.,][]{Lallement:2014, Green:2015} has recently shown promise for modeling the line-of-sight magnetic field in a small region of the nearby ($\lesssim 500$ pc) ISM \citep{vanEck:2017}. Future work incorporating both neutral and magneto-ionic data has the potential to constrain the three-dimensional magnetic field vector in three spatial dimensions in the ISM.

\acknowledgments

S.E.C. would like to thank the PSI2 program of the Universit\'e Paris-Saclay for support and hospitality during the program \textit{The ISM Beyond Three Dimensions} when work on this paper was undertaken.
We thank ISM3D organizers Marc-Antoine Miville-Desch\^enes and Josh Peek, and many other attendees for stimulating conversation, especially Vincent Guillet, Fran\c cois Boulanger, Naomi McClure-Griffiths, and Alex Hill. We also thank J. Colin Hill, Yong Zheng, Brandon Hensley, and Mary Putman for useful discussions, and the anonymous referee for thoughtful feedback. We thank Ludovic Montier for the code used to smooth noise covariance matrices. 
S.E.C. is supported by NASA through Hubble Fellowship grant \#HST-HF2-51389.001-A
awarded by the Space Telescope Science Institute, which is operated by the Association of
Universities for Research in Astronomy, Inc., for NASA, under contract NAS5-26555.

This publication utilizes data from Galactic ALFA \hi (GALFA-\HI) survey data set obtained with the Arecibo L-band Feed Array (ALFA) on the Arecibo 305m telescope. The Arecibo Observatory is operated by SRI International under a cooperative agreement with the National Science Foundation (AST-1100968), and in alliance with Ana G. M\'endez-Universidad Metropolitana, and the Universities Space Research Association. The GALFA-\hi surveys have been funded by the NSF through grants to Columbia University, the University of Wisconsin, and the University of California.

\end{document}